\documentclass[aps,prd,12pt,showpacs,preprintnumbers,groupedaddress,floatfix,nofootinbib]{revtex4}

\usepackage{amsmath}
\usepackage{graphicx}
\usepackage{color}

\def\beq{\begin{eqnarray}}
\def\eeq{\end{eqnarray}}
\def\bea{\begin{eqnarray}}
\def\eea{\end{eqnarray}}

\begin{document}

\preprint{DCP-07-03}

\title{{Fundamental and composite scalars from extra dimensions}}
\author{Alfredo Aranda}
\email{fefo@ucol.mx}
\affiliation{Facultad de Ciencias, Universidad de Colima,\\
Bernal D\'{i}az del Castillo 340, Colima, Colima, M\'exico}
\affiliation{Dual C-P Institute of High Energy Physics}

\author{J. L. D\'iaz-Cruz}
\email{lorenzo.diaz@fcfm.buap.mx}
\affiliation{Facultad de Ciencias, Universidad de Colima,\\
Bernal D\'{i}az del Castillo 340, Colima, Colima, M\'exico}
\affiliation{Facultad de Ciencias Fisico-Matem\'aticas, BUAP \\
Apdo. Postal 1364, C.P.72000 Puebla, Pue, M\'exico}
\affiliation{Dual C-P Institute of High Energy Physics}

\author{J. Hern\'andez-S\'anchez}
\email{ jaimeh@uaeh.edu.mx}  \affiliation{Centro de
Investigaci\'on en
Matem\'aticas \\
Universidad Aut\'onoma del Estado de Hidalgo, M\'exico}
\affiliation{Dual C-P Institute of High Energy Physics}

\author{R. Noriega-Papaqui}
\email{rnoriega@fisica.unam.mx} \affiliation{Instituto de F\'isica,
Universidad Nacional Aut\'onoma de M\'exico \\
Apdo. Postal 20-364, 01000 M\'exico D.F., M\'exico} 
\affiliation{Dual C-P Institute of High Energy Physics}


\begin{abstract}
\noindent We discuss a scenario consisting of an effective 4D theory
containing fundamental and composite fields. The strong dynamics
sector responsible for the compositeness is assumed to be of extra
dimensional origin. In the 4D effective theory the SM fermion and
gauge fields are taken as fundamental fields. The scalar sector
of the theory resembles a bosonic topcolor in the sense there are
two scalar Higgs fields, a composite scalar field and a fundamental
gauge-Higgs unification scalar. A detailed analysis of the scalar
spectrum is presented in order to explore the parameter space
consistent with experiment. It is found that, under the model assumptions, 
the acceptable parameter space is quite constrained. 
As a part of our phenomenological study of the model, we evaluate the
branching ratio of the lightest Higgs boson and find that our model
predicts a large FCNC mode $h \to tc$, which can be as large as O($10^{-3}$). 
Similarly, a large BR for the top FCNC decay is
obtained, namely $B.R.(t \to c+H) \simeq 10^{-4}$

\end{abstract}


\maketitle
\newpage
\section{Introduction}
\label{intro} The possible existence of extra dimensions has enabled
the creation of several physical scenarios beyond the Standard Model
(SM)~\cite{XD}. All of these scenarios contain ideas and mechanisms that
attempt to solve or to explain some of the most fundamental problems
in particle physics, i.e. the problem of electroweak symmetry
breaking (EWSB) and the problem of flavor.

It is clear by now that if extra dimensions do indeed exist, and if
they exist at the "right" scale, there is a large amount of physical
phenomena that will show up when the window is reached. Determining
the specific scenario will be a challenging task. In fact, as it
turns out, depending on which particular problem one is trying to
solve, whether it is a general problem or one generated within the
same model being proposed, there are many different possible
solutions and explanations, most of which are cleverly created to
evade experimental exclusion. This will of course change
dramatically when new experimental results come out, however it is
not clear from today's perspective how cleanly will we be able to
differentiate among the vast number of specific models.

Motivated by this situation and by the fact that there clearly are
interesting general results and mechanisms that can be drawn from
considering the possibility of extra dimensions, we take the
following approach: In order to study EWSB one can explore a 4D
effective theory of an unspecified extra dimensional theory taking
into account the effect of those general results. Following this
idea one should consider an effective theory with the possibility of
having i) a fundamental scalar field whose extra dimensional origin
is associated to a Gauge-Higgs unification scenario~\cite{Csaki:2002ur,
Scrucca:2003ut,Wulzer:2004ir,Aranda:2005ze,Sakamura:2006rf} and ii) a
composite scalar field also of extra dimensional
origin~\cite{Dobrescu:1999cs,Rius:2001dd}. We choose EWSB as a first
step and intend to do a similar analysis with flavor in future work.

This idea is similar in spirit to that of~\cite{Aranda:2000vk} and more 
recently~\cite{contino}, where it
is assumed that there is an extra dimensional strong theory that
generates heavy composite states and that at low energies the SM
fields are a possible combination of fundamental and composite
fields (except for the Higgs and $t_R$ which are required to be
purely composite fields). There is a sector associated to the
fundamental fields, one associated to the composite fields and one
corresponding to the possible mixing.

In the present proposal we consider an effective SU(2)$\times$U(1)
4D theory with a fundamental scalar and a composite scalar. In
generality the gauge bosons can be an admixture of fundamental and
composite fields as well.

\section{Scalar sector}
As discussed in the introduction we consider an effective theory
with two scalar fields $H_E$ and $H_C$. $H_E$ is a fundamental
scalar associated with components of a higher dimensional gauge
boson field as in a Gauge-Higgs unification scenario. This has the
virtue that its quartic self coupling is related to the gauge
coupling. $H_C$ on the other hand is a composite scalar with an
extra dimensional origin as well. In this case one expects the size
of the mass squared term in the potential to be of the same order of
the heavy composite states masses (assumed to be of O(TeV) in the
present work). As in~\cite{contino} we consider the following sizes
for the couplings: \beq \label{sizes} 1\sim \lambda < \lambda_c <<
4\pi \ , \eeq where $\lambda$ denotes a fundamental coupling and
$\lambda_c$ denotes the composite field coupling.

Our main interest is the scalar sector and will for the moment
assume all the gauge bosons to be fundamental fields. The Lagrangian
to be discussed is \beq \label{lagrangian} {\cal
L}_{eff}=|D_{\mu}H_E|^2+|D_{\mu}H_C|^2 -V(H_E,H_C) \ , \eeq with
\beq\label{potential}
V(H_E,H_C)=-\mu_e^2|H_E|^2-\mu_c^2|H_C|^2-\kappa^2(H_E^{\dagger}H_C+h.c.)
+\frac{g^4}{2}|H_E^{\dagger}H_E|^2+\lambda_c|H_C^{\dagger}H_C|^2 \ .
\eeq

As usually done in general Gauge-Higgs unification scenarios, the
coefficient $\mu_e$ is assumed to be radiatively generated. As
discussed above, $|\mu_c|$ is expected to be of of O(TeV), i.e. the
mass scale of composite fields in general. $\kappa$ is a free
parameter that characterizes the amount of mixing and has mass
dimension 1. Note that similar mixing parameters can appear when
more terms mixing $H_E$ and $H_C$ are incorporated into the
potential. In the present work we assume that the dominant 
contribution is the one coming from the
$\kappa$ term in Eq.~(\ref{potential}) and so we are assuming all
other mixings to be small and negligible. One interesting feature of
this Lagrangian is the form of the quartic coupling for the
fundamental scalar. Since we are relating this scalar to the gauge
structure its quartic coupling is given in terms of the SU(2) gauge
coupling $g$ as in the usual Gauge-Higgs unification scenarios.
Finally $1<\lambda_c<<4\pi$ as stated before.

Note that EWSB can be triggered by the vacuum expectation values
(vevs) of both $H_C$ and $H_E$ if $\mu_c^2$ and/or $\mu_e^2$ are
positive. In order to explore the parameters we can determine the
values of $\mu_e^2$ and $\mu_c^2$ in terms of $\kappa$, $\lambda_c$,
$g$ and the  (vevs) of the scalar fields. Denoting the vevs of $H_E$
and $H_C$ by $v_e/\sqrt{2}$ and $v_c/\sqrt{2}$ respectively,
defining $\tan\xi \equiv v_c/v_e$ and minimizing the
potential, one obtains the following expressions: \beq
\label{minimization} \nonumber \mu_e^2&=&
-\kappa^2\tan\xi+\frac{g^4}{2}v^2\cos^2\xi \, \\
\mu_c^2&=&-\kappa^2\cot\xi+\lambda_c v^2\sin^2\xi \ . \eeq

Using these expressions one can determine regions of parameter space
in $|\kappa|$ and say $\tan\xi$ using $\mu_c \sim$~TeV and
$1<\lambda_c <<4\pi$ where $\mu_e$ is positive or negative and
determine the viability of the scenario.

From the Lagrangian Eq.(\ref{lagrangian}) we determine that the
scalar, pseudoscalar and charged scalar masses are \beq
\label{masses} \nonumber  M_S^2&=&\kappa^2\left(
\begin{array}{cc}
\tan\xi+g^4v^2\cos^2\xi/\kappa^2 & -1 \\
-1 & \cot\xi+2\lambda_c v^2\sin^2\xi/\kappa^2
\end{array} \right)  \ , \\ \nonumber
M_P^2&=& M_{+}^2 = \kappa^2\left( \begin{array}{cc} \tan\xi & -1 \\
-1& \cot\xi
\end{array} \right)  \ ,
\eeq where we have defined
\beq
\label{definitions} H_E = \left( \begin{array}{c} \phi_e^+ \\
\frac{\phi_e + v_e +iA_e}{\sqrt{2}}
\end{array}\right) \ , \ \ H_C = \left( \begin{array}{c} \phi_c^+ \\
\frac{\phi_c + v_c +iA_c}{\sqrt{2}} \end{array}\right) \ . \eeq

From these expressions we obtain the physical states in the usual
way by diagonalizing the mass matrices. In the scalar sector we then
obtain the lightest ($h^0$) and heavy ($H^0$) states by performing a
rotation in the $\phi_c - \phi_e$ space parametrized by an angle
$\alpha = \alpha(\lambda_c,\xi,|\kappa|)$. Denoting the elements of
$M_S$ by $m_{ij}$, the neutral CP-even masses are

\begin{eqnarray}
m_{H, h}^2 = \frac{1}{2} \bigg(m_{11}+m_{22} \pm \sqrt{(m_{11}-m_{22})^2 + 
4 m_{12}^2} \bigg) \ ,
\end{eqnarray}
and

\begin{eqnarray}
\tan 2 \alpha = \frac{ 2 m_{12}}{m_{11}-m_{22}}  \ .
\end{eqnarray}

\begin{figure}[ht]
    \begin{center}
        \includegraphics[width=13cm]{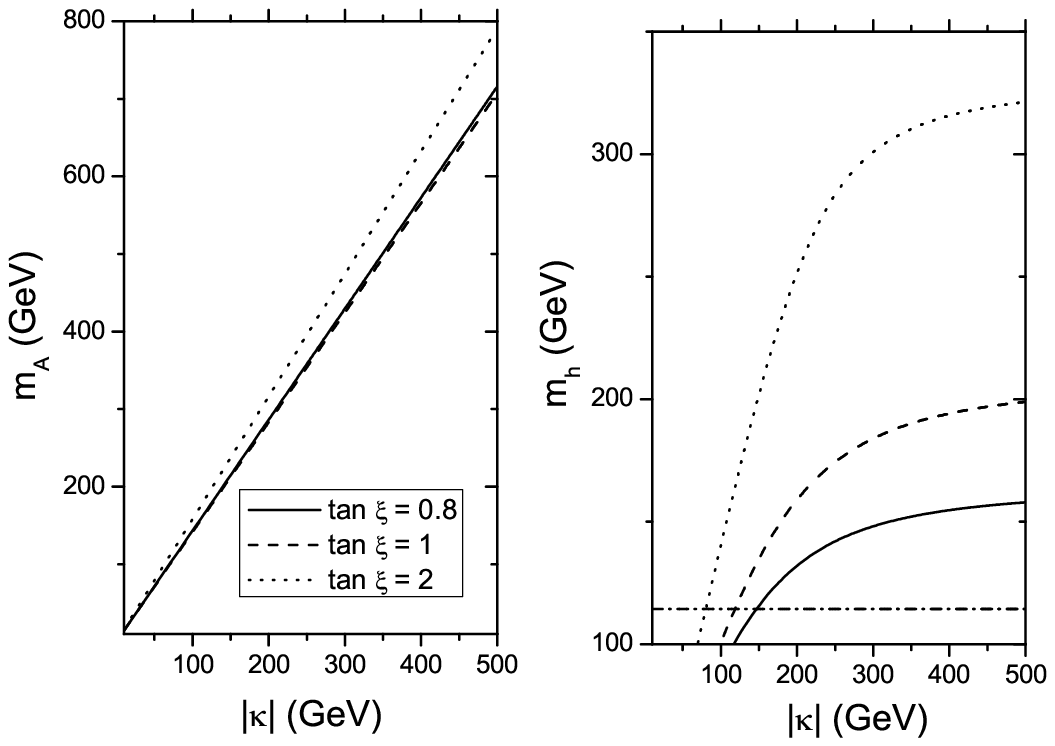}
        \includegraphics[width=13cm]{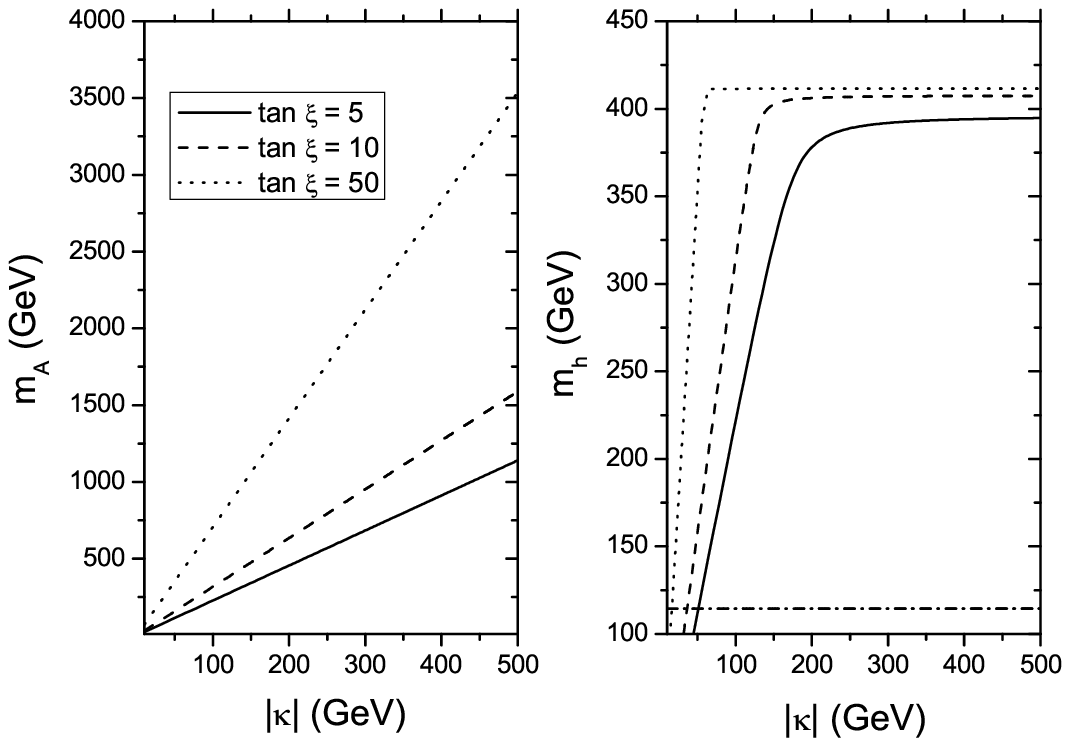}
    \end{center}
    \caption{Right plots: Pseudoscalar mass as a function of $|\kappa|$
      for different choices of $\tan\xi$. Left
      plots: The lightest scalar ($h^0$) mass as a function of
      $|\kappa|$ for the same choices of $\tan\xi$ and for $\lambda_c=1.4$.
      The horizontal line corresponds to $m_h = 114.4$~GeV.}
    \label{fig:scalarmasses}
\end{figure}

The  expression for the physical pseudoscalar mass is simple and
given by 
\beq \label{pseudomass}
    m_A^2=\kappa^2 \left( \cot\xi+\tan\xi  \right) \ .
\eeq 
In the left column of Fig.~(\ref{fig:scalarmasses}) we show the pseudoscalar mass
for different choices of $\tan\xi$ as a function of
$|\kappa|$. Note that all possibilities lie above the $\tan\xi=1$
curve and that given the form of $m_A^2$, there is a $\xi \to 1/\xi$
correspondence.

The right column of Fig.~(\ref{fig:scalarmasses}) shows
the corresponding results for the lightest scalar mass. Note that for each case
the constraint $m_h \geq 114.4$~GeV (horizontal line in the plots) sets the scale 
for $|\kappa|$. For example for
$\tan\xi = 1$,  $|\kappa| \geq 118.4$~GeV.

\begin{figure}[ht]
    \begin{center}
        \includegraphics[width=12cm]{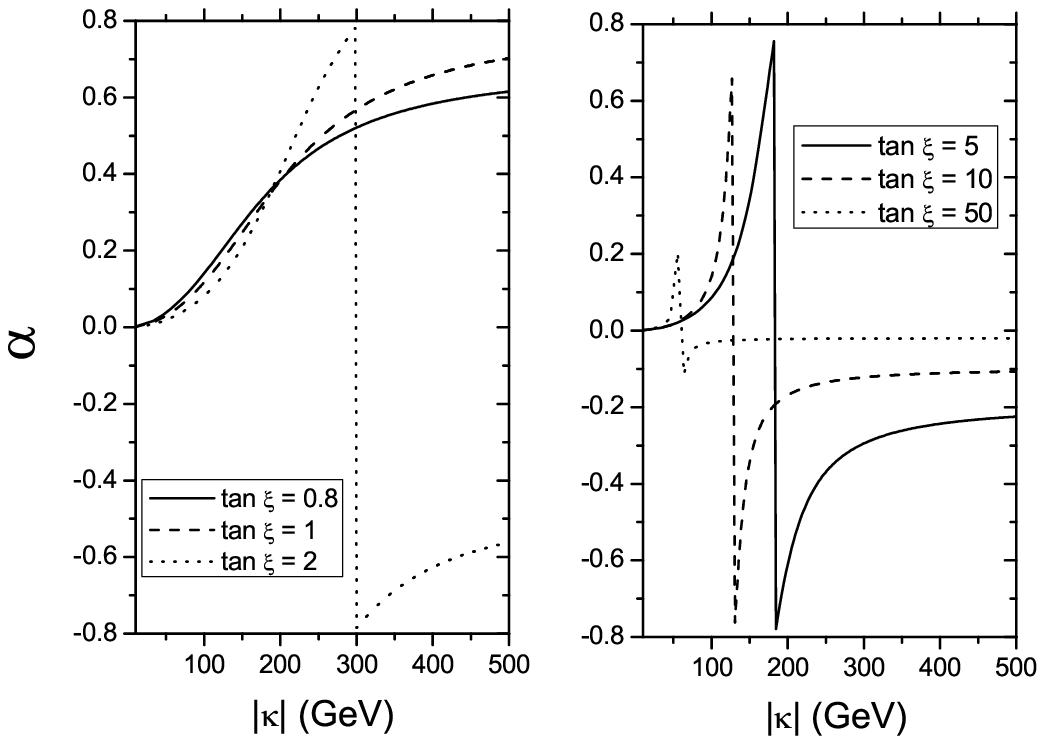}
    \end{center}
    \caption{$\alpha$ dependence on $\kappa$ for different choices of $\tan\xi$.}
    \label{fig:alpha}
\end{figure}

In order to see the explicit $\alpha$ dependence of these results, we plot it as a
function of $|\kappa|$ for the same values of $\tan\xi$ in
Fig.~(\ref{fig:alpha}).

\section{Yukawa Sector of the Model}
In this section we explore some salient aspects of the Yukawa sector
of the model presented above. For the elementary sector we consider
the types of Yukawa couplings that it could have keeping in mind
that it comes from a Gauge-Higgs unification scenario. We assume
that this elementary sector couples predominantly to the third
family, as in some particular 5D or 6D scenarios based on the gauge
group SU(3)$_w$ where the SM can be embedded~\cite{Gogoladze:2007nb,Aranda:2005st,DiazCruz:2004ss}.

For the composite sector, we use the effective Lagrangian approach
working with a scenario where the Flavor scale ($\Lambda_F$) is
assumed to be low, i.e. below the compositeness scale ($\Lambda_c$).
Then, depending on the scale, we have
\begin{itemize}
\item[i)] For $E>\Lambda_c>\Lambda_F$: A theory with elementary scalars, but
no Higgs and no Yukawas.
\item[ii)] For $\Lambda_c>E>\Lambda_F$: A Higgs, but no Yukawa operators.
\item[iii)] For $\Lambda_c \ , \Lambda_F>E$: Higgs and Yukawa operators: For
instance by using the Froggart-Nielsen mechanism we can write
\beq \label{FN}
{\cal L}_{yc}=\lambda_{ij}^u\left[\frac{\langle S\rangle}{\Lambda_f}\right]^{n_{ij}^u} \bar{Q}_i\tilde{H}
_Cu_j+\lambda_{ij}^d\left[\frac{\langle S\rangle}{\Lambda_f}\right]^{n_{ij}^d} \bar{Q}_i H_Cd_j
+ {\rm h.c.}
\eeq
\end{itemize}

Thus the Yukawa Lagrangian of our model is:
\begin{eqnarray}\nonumber
{\cal{L}}_y &=& \left[  Y^{u}_{ij} \, \bar{Q'}_{Li} \,  {\tilde H_{C}} \,  u'_{Rj} + Y^{d}_{ij} \, 
\bar{Q'}_{Li} H_{C} \, d'_{Rj} \right] \\ 
&+& \left[ \eta_t \,  \bar{Q'}_{L3} \,  {\tilde H_{E}} \, u'_{R3} 
 +   \eta_b \,  \bar{Q'}_{L3} \, H_{E} \, d'_{R3} \right] + h.c.
\label{quark_lagrangian}
\end{eqnarray}
where the first term in brackets is the contribution from the composite Higgs, the second one is 
due to the elementary Higgs which only contributes to the third family (with similar expressions 
for leptons).

After spontaneous symmetry breaking (SSB), one can derive the
quark mass matrices from Eqs.~(\ref{quark_lagrangian}) namely,
\begin{eqnarray}
\label{u_mass}
[M^u ]_{ij} &=& \frac{1}{\sqrt{2}} \left( v_{c} \, Y_{ij}^{u} + v_{e} \, \eta_t \, 
\delta_{3i} \, \delta_{3j} \right),  \\
\label{d_mass}
[M^d ]_{ij} &=& \frac{1}{\sqrt{2}} \left( v_{c} \, Y_{ij}^{d} + v_{e} \, \eta_b \, 
\delta_{3i} \, \delta_{3j} \right).
\end{eqnarray}
We now assume that the Yukawa composite matrices  $Y^u$ and $Y^d$ have the
four-Hermitic-texture form~\cite{Fritzsch:2002ga}. 
The quark mass has the same form and it is given by:
\begin{displaymath}
M_q= 
\left( \begin{array}{ccc}
0 & D_{q} & 0 \\
D_{q} & C_{q} & B_{q} \\
0 & B_{q}  & A_{q}
\end{array}\right)  \qquad 
(q = u, d) \ ,
\end{displaymath}
where $A_{q} = v_{c} \, Y_{33}^{q} + v_{e} \, \eta_{3q}$. Taking $M^u$ and  $M^d$ real,  
and following the analysis in~\cite{Fritzsch:2002ga}, we diagonalize them 
using the matrix $O$ in the following way:
\begin{equation}
\bar{M}^q = O^{T}_q \, M^{q} \, O_{q}. 
\label{masa-diagonal}
\end{equation}

From Eqs.~(\ref{u_mass}, \ref{d_mass}) we can write
\begin{equation}
[\tilde{Y}^{q}] =
\frac{\sqrt{2}}{v_c} [\bar{M}_{q}] - \eta_{q3}  \, \cot \xi \, [\tilde{h}^q] \ ,
\label{rotyukawas}
\end{equation}
where $[\tilde{Y}^{q}] =[O^{T}_q] \, [Y^{q}] \, [O_q]$, and  
$[\tilde{h}^{q}] =[O^{T}_q] \, [Diag\{ 0,\,0,\, 1\} ] \, [O_q]$.
Note that the second term (proportional to $ \eta_{q3}$) induces FCNC. Parameterizing $A_q$ as
$A_q = m_{q3} - \beta_q \, m_{2q}$~\cite{DiazCruz:2004pj} and performing the product 
$[O^{T}_q] \, [Diag\{ 0,\,0,\, 1\} ] \, [O_q]$, the term $[\tilde{h}^q] $ goes like
\begin{equation}
 [\tilde{h}^q]_{ij}  \sim \frac{\sqrt{m_{q_i} m_{q_j}}}{m_{q3}} \ .
\end{equation}

Finally the interactions of the neutral Higgs bosons $(h^{0}, H^{0}, A^{0})$ with quark pairs acquire the
following form:

\begin{eqnarray}
{\cal{L}}_Y^{q} & = &
\bar{d_{i}}\left[\frac{ m_{d_i}}{v} \frac{ \, \cos\alpha}{\sin\xi}\delta_{ij} -
\eta_b  \, \frac{\sqrt{m_{d_i} m_{d_j} } } {m_b}
\frac{ \cos(\alpha + \xi)}{\sqrt{2} \, \sin\xi}  \right]d_{j}  H^{0} \nonumber \\
&  &+
\bar{d}_{i} \left[- \frac{ m_{d_i}}{v} \frac{\sin\alpha}{\sin\xi} \delta_{ij} +  
\eta_b  \, \frac{\sqrt{m_{d_i} m_{d_j} } } {m_b} 
\frac{ \sin(\alpha + \xi)}{\sqrt{2} \, \sin\xi} \right]d_{j} h^{0} \nonumber \\
& &+i \, \bar{d}_{i}
\left[- \frac{ m_{d_i}}{v} \cot\xi \delta_{ij} + 
\eta_b  \, \frac{\sqrt{m_{d_i} m_{d_j} } } {m_b}  \frac{1 }{ \sqrt{2} \, \sin \xi} \right] 
\gamma^{5}d_{j} A^{0} \nonumber \\
& &+ 
\bar{u}_{i}\left[\frac{ m_{u_i}}{v} \frac{ \, \cos\alpha}{\sin\xi}\delta_{ij} -
\eta_t  \, \frac{\sqrt{m_{u_i} m_{u_j} } } {m_t}
\frac{ \cos(\alpha + \xi)}{\sqrt{2} \, \sin\xi} \right]u_{j}H^{0}
\nonumber \\
&  &+ \bar{u}_{i} \left[- \frac{ m_{u_i}}{v} \frac{\sin\alpha}{\sin\xi} \delta_{ij} + 
\eta_t  \, \frac{\sqrt{m_{u_i} m_{u_j} } } {m_t}
\frac{ \sin(\alpha + \xi)}{\sqrt{2} \, \sin\xi} \right]u_{j} h^{0}
\nonumber \\
& &+i \, \bar{u}_{i}
\left[ \frac{ m_{u_i}}{v} \cot\xi  \delta_{ij}  -
\eta_t  \, \frac{\sqrt{m_{u_i} m_{u_j} } } {m_t} \frac{1 }{ \sqrt{2} \, \sin \xi}  \right]
\gamma^{5} u_{j} A^{0}. 
\label{lageigenstates}
\end{eqnarray}

\section{Higgs and Top FCNC Decays}

We now present some generic numerical results in order to show the potential of this model.
A complete phenomenological study of this model is underway and will be presented elsewhere.
If we go back to the main motivation of this model, recall that we incorporated both composite 
and fundamental scalars of extra dimensional origin in order to write a two Higgs doublet model 
in 4D. Furthermore, we assumed, motivated by previous works, that the composite scalar couples 
to all quarks while the elementary (fundamental) one couples only to the third family.
Consider now the following approach: Lets suppose that naturally one would expect the top quark
to have a mass of O($m_W$) and that the reason for its heaviness is precisely that it has an
extra contribution, in this case from the elementary sector. 
Following this idea we explore the model in the case where this only happens to the top quark 
and hence set $\eta_b = 0$ in the following analysis.  

Some remarks are in order:
\begin{itemize}
\item[]  In order of get the most economical set of Yukawa parameters, we expressed the composite Yukawa 
  in terms of both the particle masses and the elementary Yukawa (see Eq.~\ref{rotyukawas}). 
  As a consequence, the fermion-fermion-CP even Higgs ($H^0$ or $h^0$) couplings have the same 
  form for both type of fermions. In the case 
  of the fermion-fermion-CP odd Higgs ($A^0$) couplings the only difference is a 
  sign in~Eq.~(\ref{lageigenstates}).
\item[] In order to avoid dangerous FCNC, we need to specify the magnitude of 
  $\eta_{3q}$. To do so, we perform the following analysis:
  The Cheng-Sher ansatz~\cite{chengsher} for the fermion-fermion-Higgs boson coupling is
\begin{equation}
\bar{f_i} \bar{f_j} \phi^0 \sim c_{ij} \frac{\sqrt{m_{f_i} m_{f_j} } }{v} \ ,
\end{equation}
and FCNC are kept under control if $| c_{ij}| \sim 10^{-1}$.
In our case the fermion-fermion-Higgs boson coupling is found to be 
\begin{equation}
\bar{f_i} \bar{f_j} \phi^0 \sim \eta_{f3} \frac{\sqrt{m_{f_i} m_{f_j} } }{m_{f_3}} \ ,
\end{equation}
and so $c_{ij} = \eta_{f3} \frac{v}{ m_{f_3}} \, \Rightarrow \,  |\eta_{f3}| = \frac{ m_{f_3} }{v} |c_{ij}|$. 
Then for u-quarks $|\eta_t | \sim 10^{-1}$ and for d-quarks $|\eta_b| \sim 10^{-2}$.
\end{itemize}

\begin{figure}[ht]
    \begin{center}
        \includegraphics[width=14cm]{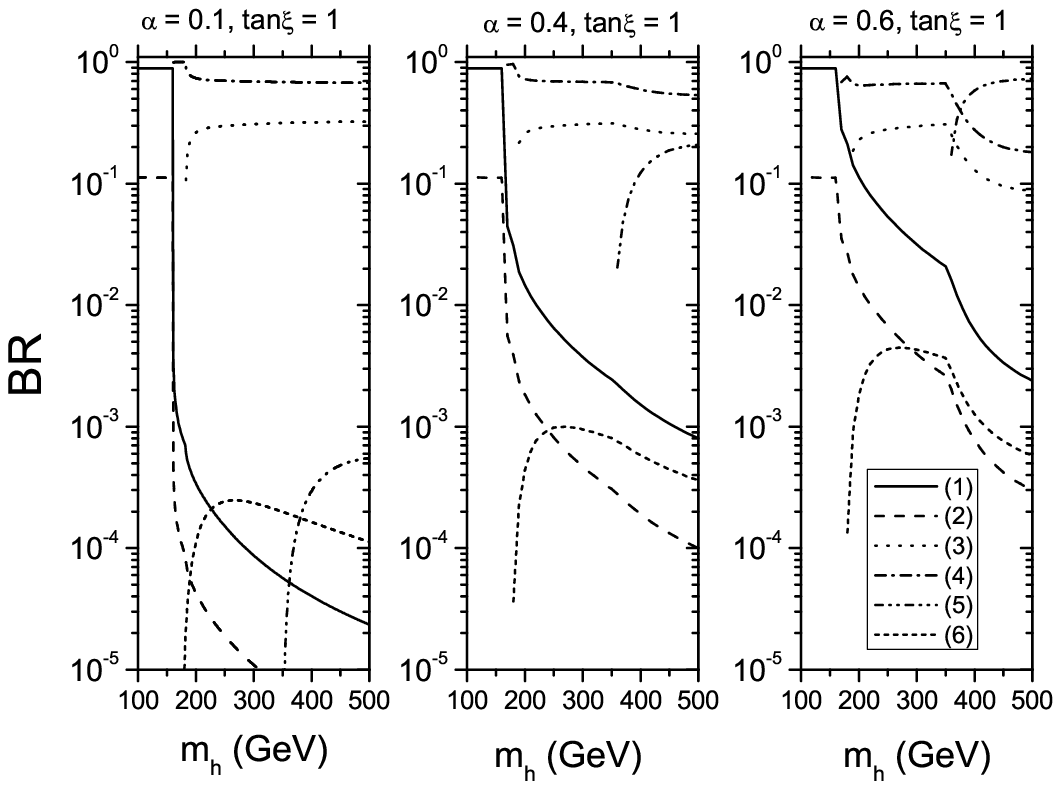}
        \includegraphics[width=14cm]{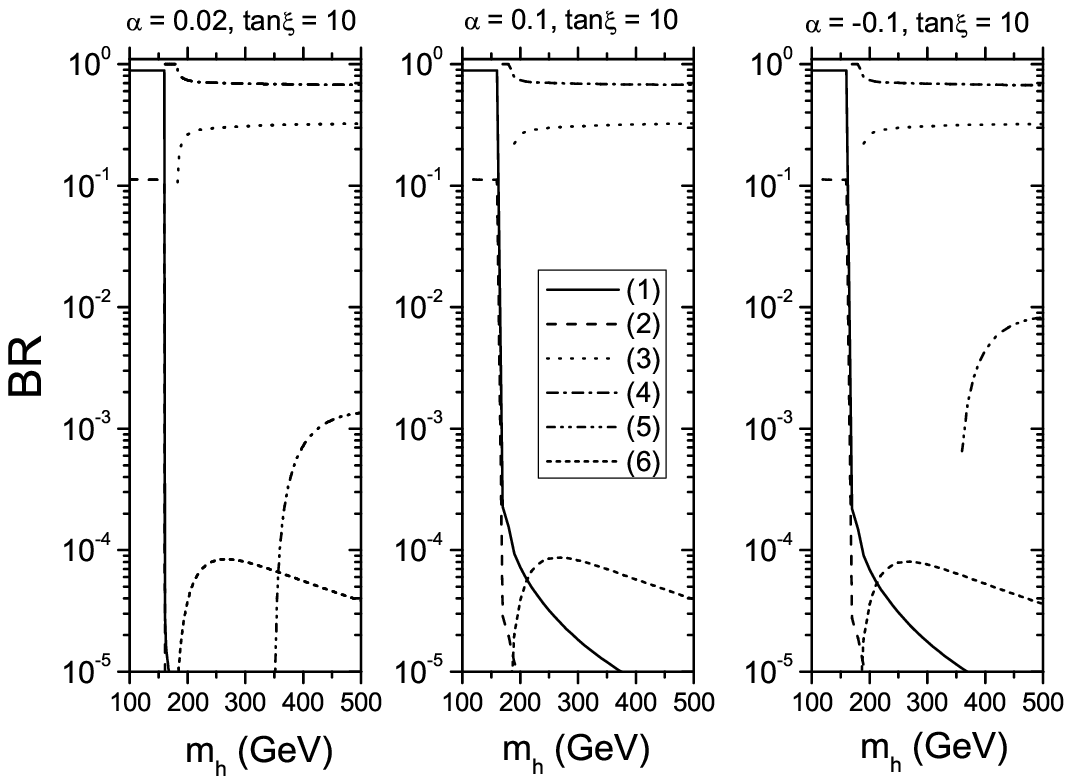}
    \end{center}
    \caption{Branching ratios for $h^0$ decay as a function of $m_h$ for $\eta_b=0$ and $|\eta_t|=10^{-1}$.
      The results are shown for $\tan\xi=1$ (top) and $\tan\xi=10$ (bottom) and for three different 
      choices of $\alpha$. These choices correspond
      to values within the range $15$~GeV $\leq |\kappa| \leq 500$~GeV. The numbers correspond to 
      the following
      modes: $b\bar{b}$ (1), $\tau \bar{\tau}$ (2), $ZZ$ (3), $WW$ (4), $t\bar{t}$ (5) and
      $t\bar{c}$ (6).}
    \label{fig:branching}
\end{figure}

Using the expressions in Eq.~(\ref{lageigenstates}) we compute the branching ratios (BR) for the 
lightest scalar $h^0$ decays to $b\bar{b}$, $\tau \bar{\tau}$, $ZZ$, $WW$, $t\bar{t}$ and
$t\bar{c}$. Fig.~(\ref{fig:branching}) shows the results for  
$\tan\xi=1$ (above) and $\tan\xi=10$ (below). We present plots for three different choices 
of $\alpha$ corresponding to 
values within the desired $|\kappa|$ 
range, i.e. $15$~GeV $\leq |\kappa| \leq 500$~GeV. Note in particular the result for the BR
corresponding to the flavor changing decay  $h^0 \to tc$. We see that for all three cases this BR 
is larger than the SM prediction by about $10$ orders of magnitude~\cite{Curiel:2003uk,Arhrib:2004xu}.

\begin{figure}[ht]
    \begin{center}
        \includegraphics[width=14cm]{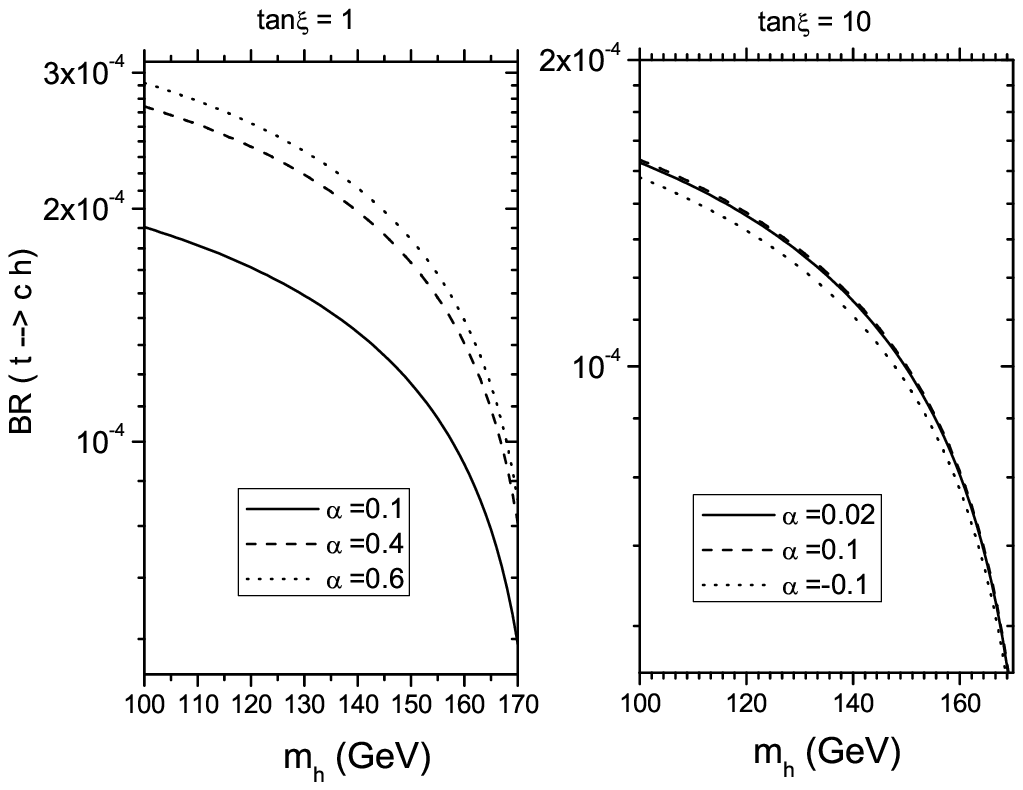}
    \end{center}
    \caption{Branching ratio for $t \to c+h$ for different choices of $\alpha$
      and $\tan\xi$.}
    \label{fig:topdecay}
\end{figure}

The discussion above deals with Higgs boson decays in our model.
However, if it so happens that the Higgs particle is light enough, then it
could show up in FCNC top decays, namely $t \to c+h$. The evaluation of
the corresponding decay branching ratio leads to
$B.R.(t\to c+h)\simeq 10^{-4}$, (see Fig.~(\ref{fig:topdecay})) which is in the 
right range to be detected at the LHC~\cite{ourtopfcnc}.

\section{Conclusions}

We have presented a model based on the fact that if there are extra dimensions of a size
relevant to particle physics phenomenology, then it is possible to have 4D scalars associated
to the extra dimensional physics. As a result, the model assumes the existence of both composite
and fundamental scalars whose origin is extra dimensional. Concretely, the model contains two
Higgs doublets, one of each kind, where one of the Higgses couples to all fermions while the
other couples only to third family fields. In this simplest version of the model all other 
fields (gauge and fermions) are treated as fundamental.

A detailed study of the scalar spectrum has revealed that this simple model has a rather 
constrained parameter space consistent with the model assumptions. 
We have computed the branching ratios
for the lightest Higgs decay and found that the one corresponding to the $h^0 \to t\bar{c}$ mode
is much larger than the one obtained in the Standard Model ($10$ orders of magnitude
approximately). Lastly we present the branching ratio
for the FCNC top decay $t \to c+h$ which we find to be of O($10^{-4}$).

\begin{acknowledgments}
JHS acknowledges support from CONACYT under grant No.50027-F and SNI. RNP acknowledges support from 
CONACYT under grant No.42026-F. AA and JLDC acnowledge support from CONACYT and SNI.
\end{acknowledgments}

\end{document}